\documentclass[pre,twocolumn]{revtex4-1}
\usepackage{amsmath}    
\usepackage{graphicx}   
\usepackage{verbatim}   
\usepackage{color}      
\usepackage{subfigure}  
\usepackage{hyperref}   
\usepackage{braket}		
\newcommand{\op}[1]{\fontdimen12\textfont3=2pt\fontdimen12\scriptfont3=1.4pt\!\null\mathop{\protect\vphantom{#1}\smash{#1}}\limits_{\sim}\null\!}

\newcommand{\xref}[1]{\protect\ref{#1}}
\newcommand{\figref}[1]{Fig.~\protect\ref{#1}}
\newcommand{\fmref}[1]{(\protect\ref{#1})}

\renewcommand{\eqref}[1]{Eq.~(\protect\ref{#1})}

\begin{document}

\title{Decoherence of a singlet-triplet superposition state
  \\ under dipolar interactions of an uncorrelated environment}
\author{Patrick Vorndamme and J\"urgen Schnack}
\affiliation{ Department of Physics, Bielefeld University, 33615 Bielefeld, Germany}

\date{\today}

\begin{abstract}
Recently, it was shown that by means of an scanning tunneling 
microscope it is experimentally possible to stimulate clock transitions 
between the singlet and the non-magnetic triplet state of a Heisenberg-coupled 
spin dimer [Bae et al., Sci. Adv. 4, eaau4159 (2018)]. This leads to more strongly 
protected clock transitions while ordinary ones only provide first-order protection 
against magnetic noise. However, large decoherence times of clocklike states normally 
refer to ensembles of spins which do not dephase. In the cited experiment, only one
single dimer is manipulated and not an ensemble. For this reason, we simulate how a 
single dimer behaves in an environment of other spins which couple to the dimer via 
dipolar interactions. We perform unitary time evolutions in the complete Hilbert space, 
including dimer and a reasonably large environment. We will see that for a weak environment, 
this approach  confirms long decoherence times for the clocklike state, but with stronger 
couplings this statement does not hold. As a reference, we compare the behavior of the dimer 
with other, non-clocklike, superposition states. Furthermore, we show that the internal 
dynamics of the bath plays an important role for the decoherence time of the system. 
In a regime where the system is weakly coupled to the bath, stronger interactions among 
environmental spins worsen the decoherence time up to a certain degree, while if system and 
bath are strongly coupled, stronger interactions in the environment improve decoherence times. 

\end{abstract}

\maketitle

\section{Introduction}

To perform quantum computing, it is necessary to have
building blocks that are individually controllable and whose
superposition states have long decoherence times. Spin systems
that exhibit clock transitions are promising candidates at
least for the last property \citep{Baeeaau4159,PhysRevB.100.064405,Shiddiq2016,
  Wolfowicz2013}. Such clock transitions mean a stimulation
between two energy eigenstates $\ket{E_1}$ and  $\ket{E_2}$ of
the system which are independent of the external magnetic field
at least to first order \citep{PhysRevB.100.064405}. This leads
to a precession of the phase difference in a superposition of
these two states with a frequency $\omega = E_1 - E_2$ which is
thus also independent of the external magnetic field. When an
experimenter excites an ensemble of such systems, all spins
precess at the same frequency $\omega$ and do not dephase, regardless of local
magnetic field fluctuations. Experimentally, this results in
large $T_2^{*}$ times.

In this paper we investigate the decoherence
behavior of a single spin dimer which is dipolar coupled to a 
bath of environmental spins, motivated by the experiment
described in \citep{Baeeaau4159}. Regarding decoherence this is
a completely different scenario compared to dephasing of an ensemble of
spins. 
As an approximation of the exact environment in the experiment
we use a model system in 
which the environmental spins are randomly distributed on a
spherical surface around the spins of the dimer. Therefore, the
absolute values of our decoherence times are not realistic, but
we can make relative statements in the sense that scenario A has
a much longer decoherence time than scenario B, using our
environment as a test bed.

Generally, local magnetic field fluctuations can have many
sources. Beyond an inhomogeneity in the external field, referred
to as non-intrinsic decoherence, intrinsic effects such as
dipolar 
interactions of near nuclei are important \citep{RBM:PRB12}. But not only that,
for molecular spin clusters it was shown that interactions with
neighboring electronic spins play an important role, too
\citep{LUNGHI2019165325}. Thus, we investigate in our model both 
cases: strong and weak magnetic moments of the 
environmental spins.

Figure~\xref{fig:energie0} illustrates the system we are interested
in, a Heisenberg coupled spin dimer with $s_1 = s_2 = 1/2$ and thus four energy
levels. Two of them are completely independent of the external
magnetic field, not only to first order. We will refer to a
superposition between these two states as our clock like
scenario.   Experimentally, it is possible to create and
manipulate these superpositions \citep{Baeeaau4159}. In view of
this, such systems are individually controllable; this meets
an important criterion for the
usability in the context of quantum computing as pointed out in
the beginning. In the experiment
the dimer consists of two titanium atoms, a localized
time-dependent magnetic field was realized by means of an STM
tip. The tip moved one atom in an inhomogeneous magnetic
field. In this way, the atom experienced a time-dependent
magnetic field. A small difference in the Land\'{e} factors of the
two atoms was also compensated by the tip.

As a side note we want to point out that manipulation of single
(molecular) spins is usually difficult. Magnetically this is possible by
means of an STM. But in the area of spintronics much research is
also being devoted to how individual spins can be manipulated by
means of time-dependent 
electric fields as an alternative approach \citep{PhysRevLett.122.037202, GFB:PRL17}.

The paper is organized as follows. In Section \ref{sec-2} we
introduce the theoretical model.
In Section~\ref{sec-3} we explain the different scenarios we
investigate for the initial state of the dimer and how we
prepare the environment. We also point out why decoherence is a
process for which no energy exchange between system and
environment is needed; it therefore differs from relaxation and
thermalization. In Section~\ref{sec-4} we show our numerical
results.  
The article closes with a discussion in Section~\ref{sec-5}.

\section{Model}
\label{sec-2}

The Hamiltonian of our model consists of three parts 
\begin{equation}
  \op H = \op H_S + \op H_{SE} + \op H_E
\ .
\label{hamilton}
\end{equation}

The first part $\op H_S$ (system Hamiltonian) describes the spin
dimer and contains Heisenberg and Zeeman terms 
\begin{equation}
\op{H}_S = J \op{\vec{s}}_1 \cdot \op{\vec{s}}_2 + g_S \mu_S B
(\op{s}_1^z + \op{s}_2^z)
\ .
\end{equation}

\begin{figure}[h!]
\centering
\includegraphics*[clip,width=0.85\columnwidth]{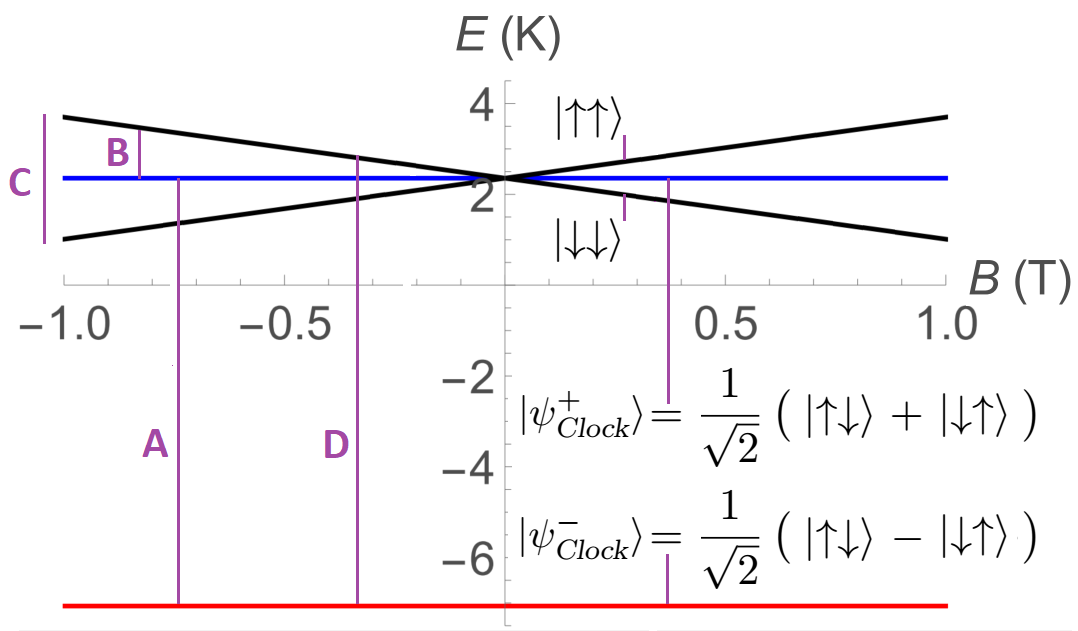}
\caption{\label{fig:energie0} Energy eigenvalues of
  $\op{H}_S$. The singlet state is shown in red, the
  non-magnetic triplet state in blue and the other two triplet
  states in black.}
\end{figure}

The magnetic field $B$ is constant and points into
$z$-direction. The coupling constant $J$ is chosen to be
antiferromagnetic ($J = 9.425$ K, with $\hbar = k_B =
1$) and of the same order of magnitude as   measured in the
experiment \citep{Baeeaau4159}. The magnetic interaction
strength $g_S \mu_S = 1.3434$ K/T is chosen to be the same as
for free electrons.  The dimer consists of two spins $s_1 = s_2 = 1/2$ so
that $\op{H}_S$ has four energy eigenvalues. The spins couple
either to total spin $S = 0$ (singlet), or to a spin
$S = 1$ (triplet). As marked in \figref{fig:energie0},
the eigenstates of the singlet and the 
non-magnetic triplet state are given by
\begin{equation}
\ket{\psi_{\textsl{Clock}}^{\pm}} = \frac{1}{\sqrt{2}} \left( \ket{\uparrow
  \downarrow} \pm \ket{\downarrow \uparrow} \right)
\ ,
\label{eigenstates1}
\end{equation}

and the other two states of the triplet are provided by the
polarized states $\ket{\uparrow \uparrow}$ and $\ket{\downarrow
  \downarrow}$.

The second part $\op H_{SE}$ of Hamiltonian \eqref{hamilton}
contains dipolar interactions between dimer and $(N-2)$
environmental spins 
\begin{equation}
\op{H}_{SE} = \sum_{i = 1}^2 \sum_{j = 3}^{N}
\frac{A_1}{r_{ij}^3} \left( \op{\vec{s}}_i  \cdot \op{\vec{s}}_j
-  \frac{3 (\op{\vec{s}}_i  \cdot \vec{r}_{ij}  ) (
  \op{\vec{s}}_j \cdot \vec{r}_{ij} )}{r_{ij}^2} \right) 
\end{equation}

with constant
\begin{equation}
  A_1 = \frac{\mu_0 g_S \mu_S g \mu}{4 \pi}
\ .
\end{equation}

We will use the factor $g \mu$ as tunable parameter for the
dipolar interaction strength of the environmental spins and
therefore the coupling between system and environment. 

The last part $\op H_{E}$ contains dipolar interactions
between different environmental spins and their Zeeman terms 
\begin{align}
\op{H}_E =& \lambda \sum_{i = 3}^{N} \sum_{j = i +1}^{N}
\frac{A_2}{r_{ij}^3} \left( \op{\vec{s}}_{i} \cdot
\op{\vec{s}}_{j} -  \frac{ 3 ( \op{\vec{s}}_{i} \cdot
  \vec{r}_{ij}  ) ( \op{\vec{s}}_{j} \cdot \vec{r}_{ij}
  )}{r_{ij}^2} \right) \notag \\  
&+ \sum_{i = 3}^{N} g \mu (B + \Delta B_i) \op{s}_i^z
\label{henv}
\end{align}
 
with constant 
\begin{equation}
A_2 = \frac{\mu_0 (g \mu)^2}{4 \pi}
\end{equation}

and magnetic fluctuations $\Delta B_i$ at the individual
positions of the environmental spins. We found that in our
scenarios these inhomogeneities make no difference; we therefore
apply in the following $\Delta B_i = 0$ $\forall i$. The
factor 
$\lambda$ in Hamiltonian \eqref{henv} allows us to scale the
dipolar interactions among environmental spins only,
without changing $\op H_{SE}$. Increasing the
value of $\lambda$ is comparable 
to a situation where the environmental spins are closer together
and therefore interact stronger.

\begin{figure}[h!]
\centering
\includegraphics*[clip,width=0.85\columnwidth]{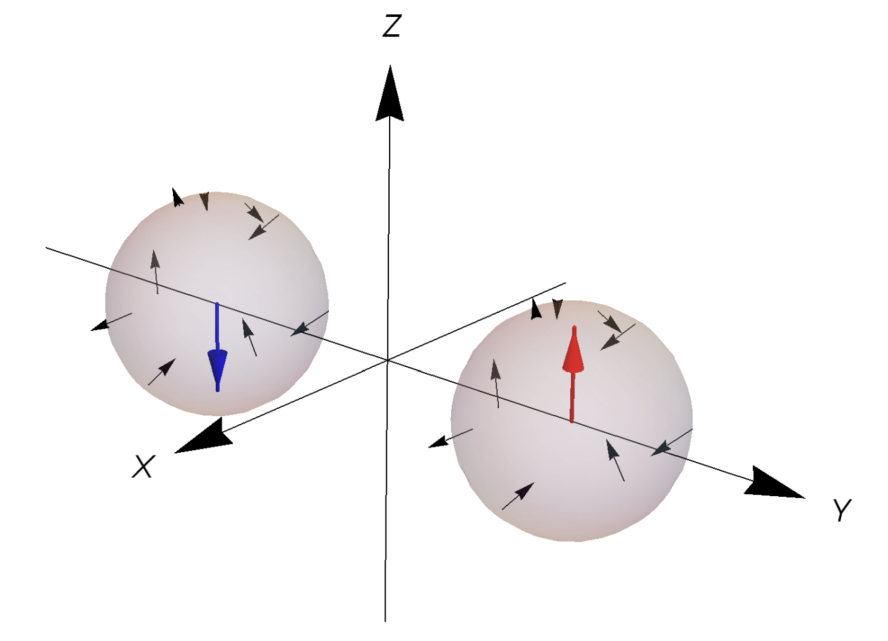}
\caption{\label{fig:2} Symbolic visualization of the
  investigated model. The two spins of the dimer are shown in
  red and blue and there are nine environmental spins on a
  spherical surface around both of them.}
\end{figure}

In case of dipolar interactions we need to choose spatial
coordinates of all spins. Altogether we choose $N = 20$ spins of
which $(N -2)$ are environmental spins. We arrange each half of
them randomly around the two spins of the dimer on a spherical
surface with radius $R = 1.5 \; \mathring{A} $. The model is
illustrated in Figure~\xref{fig:2}.

We pretend the two spins of the dimer $s_1$ and $s_2$ far apart
($r_{ij} \rightarrow \infty$), so that with respect to the
dipolar interactions the two clusters of environmental spins do
not mutually interact. This simplifies the calculation and
should not be unphysical, since the dipolar interactions
decrease with $r_{ij}^3$. This trick also allows us to
diagonalize the Hamiltonian $\op H_S + \op H_E$ (without $\op
H_{SE}$) because the effective Hilbert space is smaller and we
can show the energy spectrum of system and environment. But in
order to perform time evolution including $\op H_{SE}$ the
Hilbert space is too large to diagonalize the Hamiltonian, and we
rely on other numerical methods. 

\begin{figure}[h!]
\centering
\includegraphics*[clip,width=0.75\columnwidth]{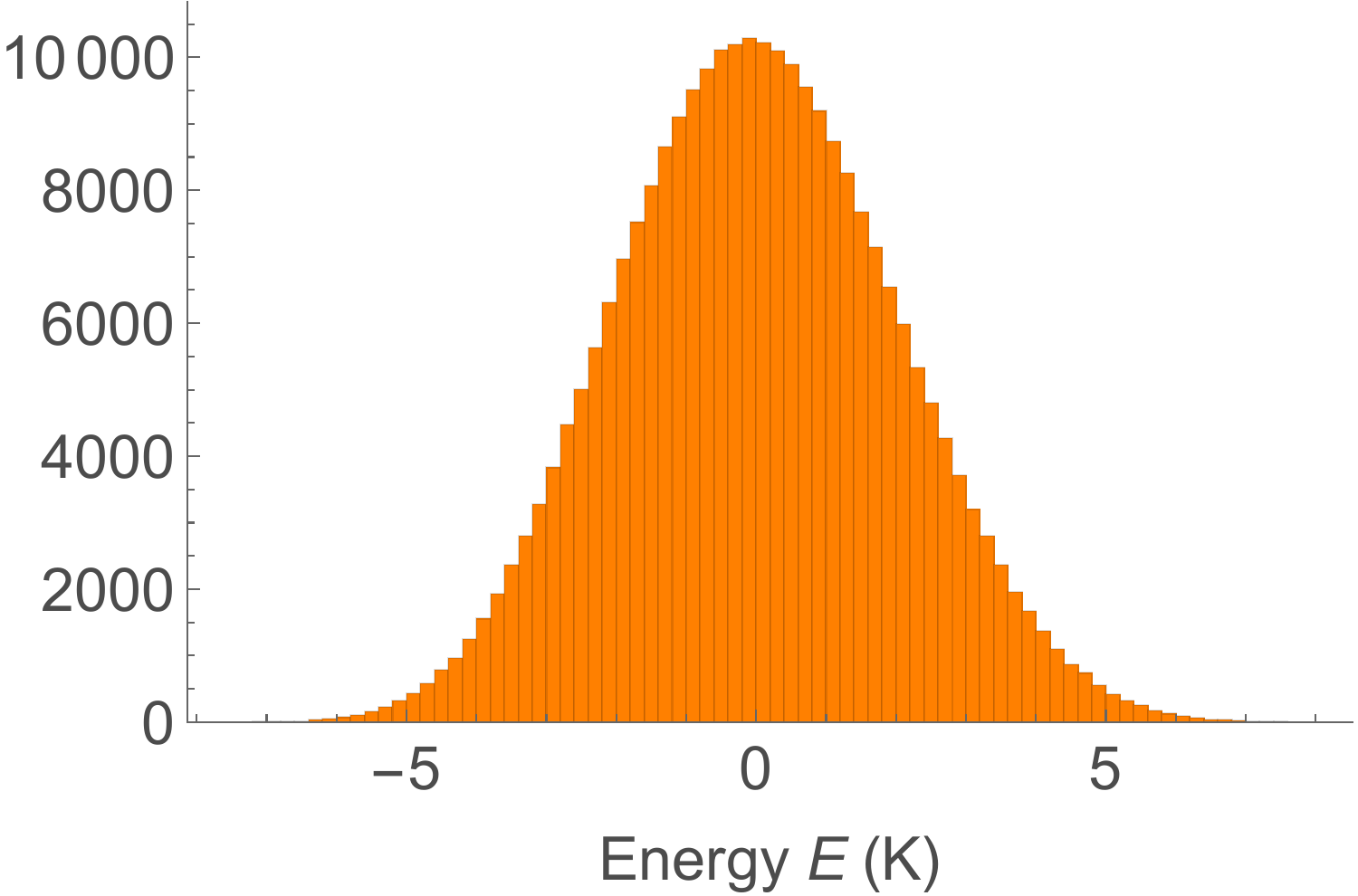}
\caption{\label{fig:3} Histogrammed number of energy eigenvalues of
  environmental Hamiltonian $\op H_E$ at $B = -1$ T.
  The magnetic interaction strength is chosen
  to be $g \mu = 0.6717$ K/T with scaling parameter $\lambda =
  1$. The total number of energy eigenvalues is $2^{18}$.} 
\end{figure}

\begin{figure}[h!]
\centering
\includegraphics*[clip,width=0.75\columnwidth]{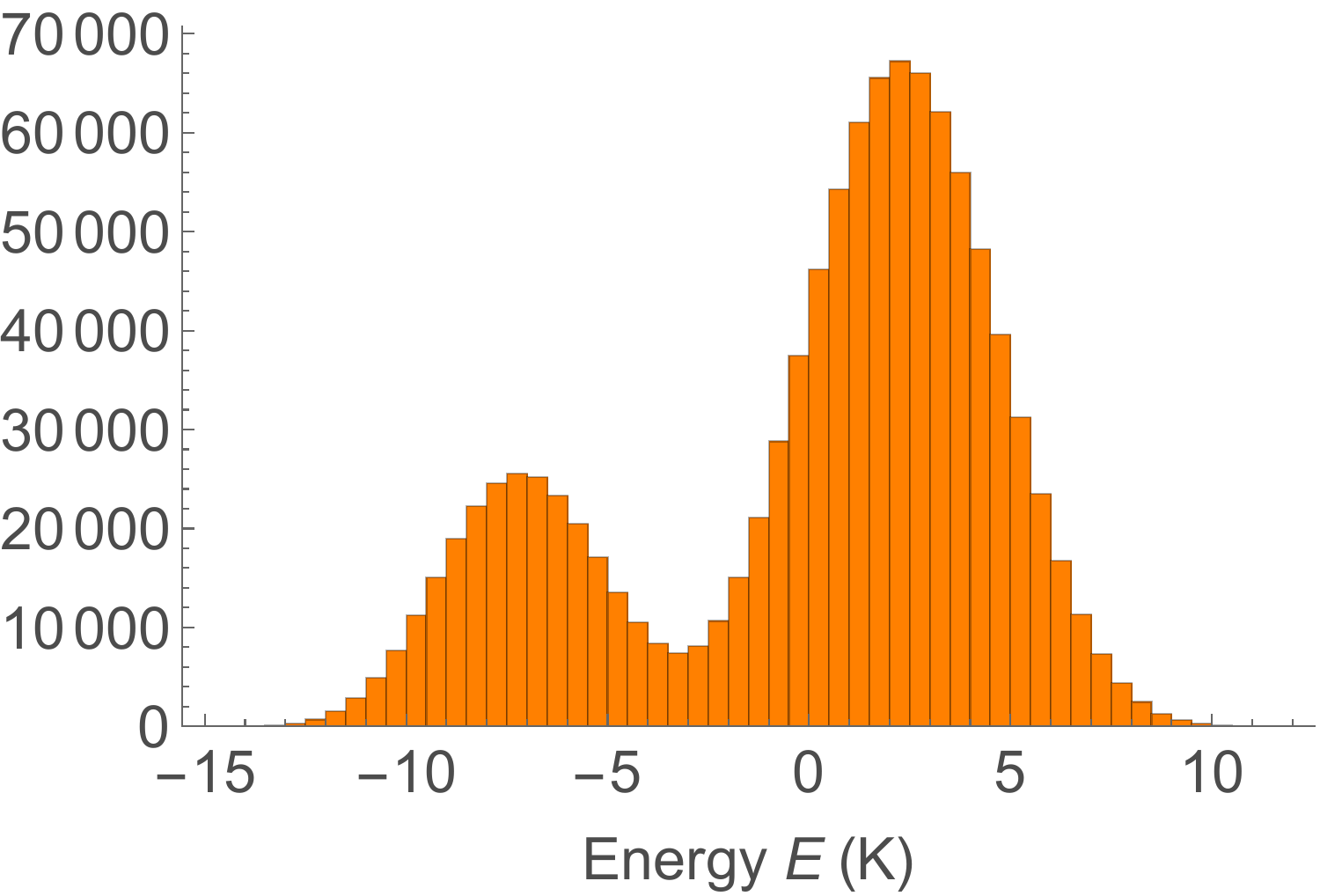}
\caption{\label{fig:4} Histogrammed number of energy eigenvalues of
  Hamiltonian $\op H_S + \op H_E$ at $B = -1$ T.
  The magnetic interaction strength is chosen
  to be $g \mu = 0.6717$ K/T with scaling parameter $\lambda =
  1$. The total number of energy eigenvalues is $2^{20}$.} 
\end{figure}

Figure~\xref{fig:3} shows the energy spectrum of $\op H_E$ with
parameters $B = -1$ T, $g\mu = 0.6717$ K/T and $\lambda =
1$. Figure~\xref{fig:4} shows the combined spectrum of $\op H_S +
\op H_E$ with the same parameters. Most energy eigenvalues are
centered around the singlet and the triplet region of the dimer,
which gives two peaks in the distribution of energy values but also energies
in between. Therefore, the
interaction $\op H_{SE}$ will cause transitions between levels of this
spectrum when the system is time-evolved with the full
Hamiltonian \fmref{hamilton}, even for small interactions $\op H_{SE}$.

In general, our approach is very similar to central spin models
with the difference that we have two spins of interest in the
center \citep{PhysRevB.88.155305, PhysRevLett.100.160505}.

\section{Preparations}
\label{sec-3}

In all following calculations we initialize the total state in a product
\begin{equation}
\ket{\psi (t = 0)} = \ket{\psi_S} \otimes \ket{\psi_E}
\label{initial}
\end{equation}

of dimer and its environment. As initial state of the dimer we
investigate different scenarios. We choose between four
initial states $\ket{\psi_S}$ that are all
superpositions of eigenstates  of $\op{H}_S$  
\begin{align}
\label{scena1}
\textsf{A:} \;& \ket{\psi_S} = \frac{1}{\sqrt{2}} \left( \ket{\psi_{\textsl{Clock}}^+} +  \ket{\psi_{\textsl{Clock}}^-} \right) = \ket{\uparrow \downarrow}  \\
\label{scena2}
\textsf{B:} \;& \ket{\psi_S} = \frac{1}{\sqrt{2}} \left( \ket{\psi_{\textsl{Clock}}^+} + \ket{\downarrow \downarrow} \right)  \\
&\; \; \; \; \; \; \; \;= \frac{1}{2} \ket{\uparrow \downarrow} + \frac{1}{2} \ket{\downarrow \uparrow} + \frac{1}{\sqrt{2}} \ket{\downarrow \downarrow} \notag \\
\label{scena3}
\textsf{C:} \;& \ket{\psi_S} = \frac{1}{\sqrt{2}} \left( \ket{\uparrow \uparrow} + \ket{\downarrow \downarrow} \right)  \\
\label{scena4}
\textsf{D:} \;& \ket{\psi_S} = \frac{1}{\sqrt{2}} \left( \ket{\psi_{\textsl{Clock}}^-} + \ket{\downarrow \downarrow} \right)  \\
&\; \; \; \; \; \; \; \;= \frac{1}{2} \ket{\uparrow \downarrow} - \frac{1}{2} \ket{\downarrow \uparrow} + \frac{1}{\sqrt{2}} \ket{\downarrow \downarrow}.\notag
\end{align}

To make a statement if the initial state of the dimer
in scenario A has a long decoherence time we need to compare it
with a reference in the same model environment. Our reference
will be the scenarios B, C and D which are non-clocklike
superpositions. Due to dipolar interactions with the
environmental spins, the state \eqref{initial} will not remain a
product state: system and environment will entangle.

Since our Hamiltonian is time-independent, time evolution is given by 
\begin{equation}
\ket{\psi (t)} = \op U(t) \ket{\psi (0)} = e^{-i \op H t} \ket{\psi (0)}
\end{equation}
with time evolution operator $\op U(t)$ and full
Hamiltonian $\op H$, \eqref{hamilton}. We calculate the evolution
with a Suzuki-Trotter product expansion numerically exact
\citep{raedt2004computational}.
As the initial state of the
environment $\ket{\psi_E}$ we choose a random state with
Gaussian distributed coefficients, both for real and
imaginary parts. In this way, we reach all of the possible states
with the same probability
\cite{Haa:AM33,CoS:CMP06,BaG:PRL09,Rei:NC16}. The state
$\ket{\psi_E}$ 
is maximally uncorrelated, also referred to as an infinite
temperature state.

In general, the state of the environment can have a considerable impact
on the decoherence behavior of 
the system as already pointed out in
\citep{PhysRevB.77.184301, PhysRevA.87.022117}. For example the
environment could be at a lower temperature. In such a case, not
the full energy spectrum of the environment is occupied
if the width of the
spectrum is much larger than $k_B T$, which changes the
thermalization (and maybe also decoherence) process
\citep{PhysRevE.88.042121}. This is typically not the case for nuclear
spins with small magnetic moments, but in the case of a dense
electronic environment the temperature of the latter could become
important. Such effects will not be covered in this
paper.

Regarding the set of all possible environmental states
low temperatures are special. The overwhelming majority of all
possible states we obtain by choosing a random state will be
close to infinite temperature and behave the same, i.e. typically
according to the concept of typicality \cite{GMM:09,BRG:18,ReG:PA19}.
For this reason the dynamics we will
show for one single random state
already represents the dynamics for most of all possible environmental
states, cf. \cite{PhysRevE.97.062129,BaG:PRL09,Rei:NC16,PhysRevLett.122.080603}. 

All information about the dimer is contained in the reduced density matrix 
\begin{equation}
\op \rho = \textsl{Tr}_{E} \left( \op{\rho}_{SE} \right),
\label{density}
\end{equation}
in which $\op{\rho}_{SE} = \ket{\psi} \bra{\psi}$ is the density
matrix of the quantum state of the total system. Regarding the
initial state \eqref{initial} the reduced density matrix
\eqref{density} describes a pure state, but becomes mixed over
time through interactions $\op{H}_{SE}$. This process of
entanglement is the essence of decoherence for an observer of
the dimer \citep{RevModPhys.76.1267,
  2014arXiv1404.2635S}. Written in the basis of the eigenstates
of $\op{H}_S$, the reduced density matrix \eqref{density}
contains non-diagonal interference terms which decay over
time.

There are various ways of quantifying decoherence; for
example  purity $\textsl{Tr} \left( \op{\rho}^2 \right)$ or the
von Neumann entropy $S = - \textsl{Tr} \left( \op \rho \ln \op
\rho \right)$ \citep{SciPostPhys.2.2.010}. We decided to look
directly at the relevant (depending on the scenario)
non-diagonal elements of the reduced density matrix. The more
entangled the system and environment are, the smaller the
absolute value of these matrix elements becomes and the more quantum mechanical 
superpositions of the dimer are destroyed. Of course,
superpositions and coherence still exist at the level of the complete
state of the total system including
the environment, but experimentally the measurement statistics
of the dimer as a subsystem turns more and more into a classical
mixture. 

We want to point out that this entanglement and therefore the
decay of the non-diagonal elements of the reduced density matrix
in general do not require a substantial energy exchange between system
and environment \citep{Obada_2007}. Imagine a product state such as
\eqref{initial} and the system in a superposition $\ket{\psi_S}
= \ket{\psi_{S_1}} + \ket{\psi_{S_2}}$. The state
\eqref{initial} can then be written as 
\begin{align}
\ket{\psi} &= \frac{1}{\sqrt{2}} \left( \ket{\psi_{S_1}} + \ket{\psi_{S_2}} \right) \otimes \ket{\psi_E} \notag \\
&= \frac{1}{\sqrt{2}} \ket{\psi_{S_1}} \otimes \ket{\psi_E} + \frac{1}{\sqrt{2}} \ket{\psi_{S_2}} \otimes \ket{\psi_E}.
\label{example}
\end{align} 

If the Hilbert space of the environment is very large, there
will exist states $\ket{\psi_{E^{'}}}$ which lie infinitesimally
close in energy but are orthogonal to $\ket{\psi_E}$,
$\braket{\psi_E | \psi_{E^{'}}} = 0$. If the interaction between
system and environment propagates the state \eqref{example} into  
\begin{align}
  \rightarrow \;\;\; \frac{1}{\sqrt{2}} \ket{\psi_{S_1}} \otimes \ket{\psi_E}
  + \frac{1}{\sqrt{2}} \ket{\psi_{S_2}} \otimes \ket{\psi_{E^{'}}}
\label{example2}
\end{align} 

the energy distribution between system and environment has not
changed, but the non-diagonal elements in the reduced density
matrix that represent the superposition of $\ket{\psi_{S_1}}$
and $\ket{\psi_{S_2}}$ have completely decayed.

\section{Calculations}
\label{sec-4}

In all following calculations the external magnetic field is
fixed to be $B = -1$ T. We investigate the behavior of the four
different initial scenarios A, B, C and D described by equations
\fmref{scena1}, \fmref{scena2}, \fmref{scena3} and
\fmref{scena4}. To begin with, we fix $\lambda = 1$ and vary the
magnetic interaction strength $g \mu$ of the environmental
spins, which affects both $\op H_{SE}$ and $\op H_{E}$. 

\begin{figure}[ht!]
\centering
\includegraphics*[clip,width=0.75\columnwidth]{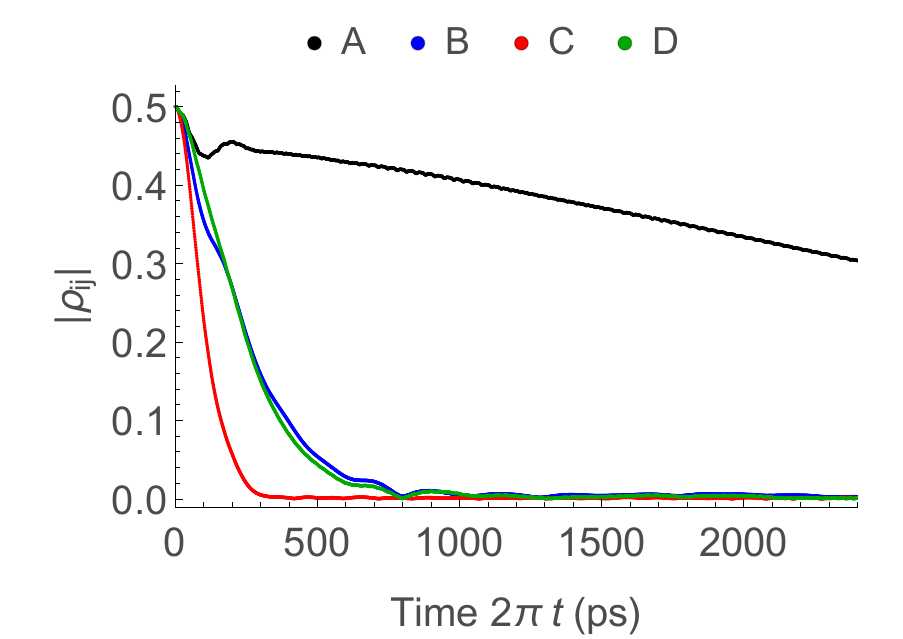}
\caption{\label{fig:5} Evolution of the absolute value of
  the relevant non-diagonal element $|\rho_{ij}|$ of the reduced
  density matrix \eqref{density} in the different scenarios A, B, C and D described in equations \fmref{scena1}, \fmref{scena2}, \fmref{scena3} and \fmref{scena4}. The chosen
  parameters are $\lambda = 1$, $g \mu = 0.3359$ K/T and $B =
  -1$ T.} 
\end{figure}

\begin{figure}[ht!]
\centering
\includegraphics*[clip,width=0.75\columnwidth]{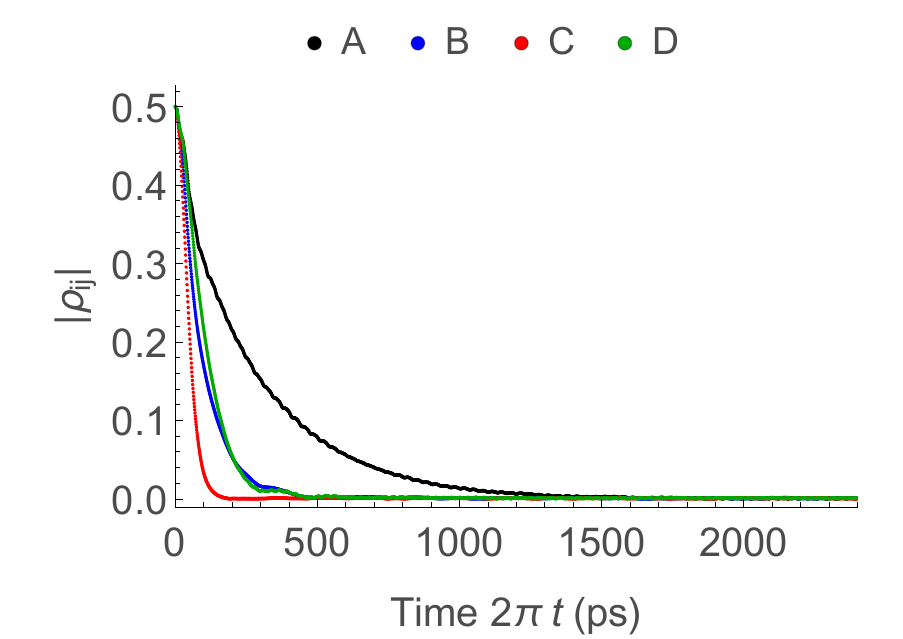}
\caption{\label{fig:6} Same as \figref{fig:5}, but with $g \mu = 0.6717$ K/T.}
\end{figure}

\begin{figure}[ht!]
\centering
\includegraphics*[clip,width=0.75\columnwidth]{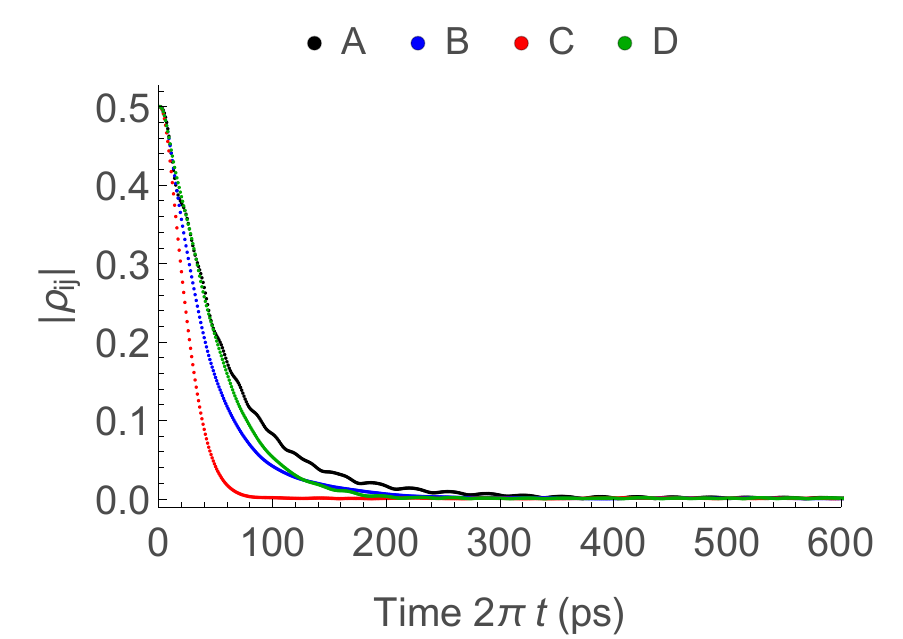}
\caption{\label{fig:7} Same as \figref{fig:5}, but with $g \mu = 1.3434$ K/T and a shorter period of time.}
\end{figure}

In Figs. \xref{fig:5}, \xref{fig:6} and \xref{fig:7} the
absolute 
value of the relevant non-diagonal element
$|\rho_{ij}|$ of the reduced density matrix \eqref{density} is
shown. For scenario A this is $|\bra{\psi_{\textsl{Clock}}^+} \op \rho
\ket{\psi_{\textsl{Clock}}^-}|$, for scenario B it is $|\bra{\psi_{\textsl{Clock}}^+} \op
\rho \ket{\downarrow \downarrow}|$, for scenario C it is $|\bra{\uparrow \uparrow} \op
\rho \ket{\downarrow \downarrow}|$ and for scenario D it is $|\bra{\psi_{\textsl{Clock}}^-} \op
\rho \ket{\downarrow \downarrow}|$. These values decay
through interactions with the environment $\op H_{SE}$, in most
of the shown cases approximately exponentially. But for the weaker
environments a Gaussian like decay is possible as also
pointed out in \citep{PhysRevB.77.184301}. 

\begin{figure}[ht!]
\centering
\includegraphics*[clip,width=0.75\columnwidth]{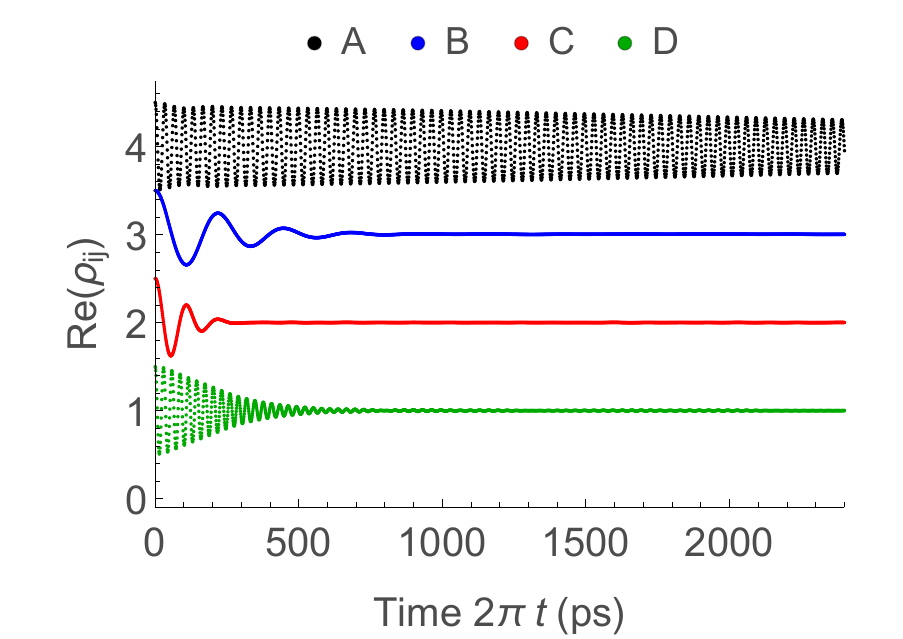}
\caption{\label{fig:8} Same as \figref{fig:5}, but the real
  part of $\rho_{ij}$ (with integer offsets) is shown instead
  of the absolute value. The oscillation frequency in the
  different scenarios A, B, C and D is given by the transition
  energies in Fig. \ref{fig:energie0}. The amplitude of this oscillation is given by the absolute value of $\rho_{ij}$.} 
\end{figure}

\begin{figure}[ht!]
\centering
\includegraphics*[clip,width=0.75\columnwidth]{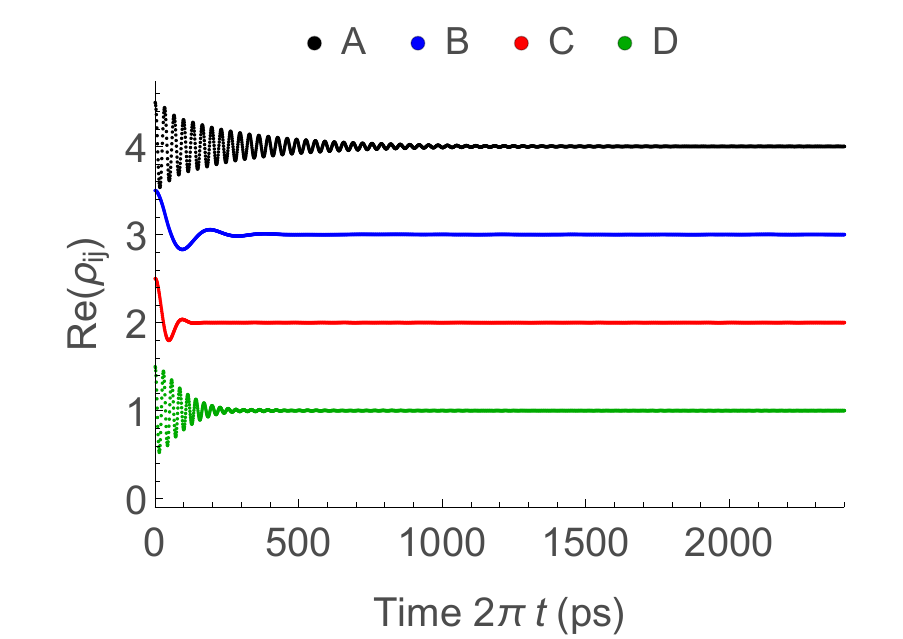}
\caption{\label{fig:9} Same as \figref{fig:8}, but with $g \mu = 0.6717$ K/T.}
\end{figure}

\begin{figure}[ht!]
\centering
\includegraphics*[clip,width=0.75\columnwidth]{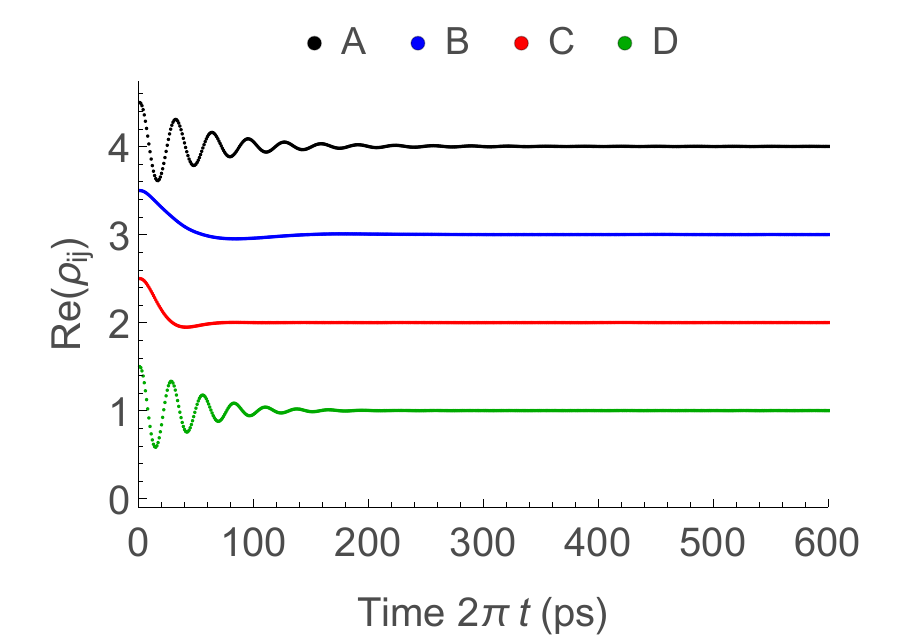}
\caption{\label{fig:10} Same as \figref{fig:8}, but with $g \mu = 1.3434$ K/T and a shorter period of time.}
\end{figure}

Figs. \xref{fig:8}, \xref{fig:9} and \xref{fig:10} show the
associated real parts of these matrix elements. They oscillate 
with  a frequency $\omega$ equal to the transition
energy of the two $\op H_S$ eigenstates the superposition is
built of (cf. \figref{fig:energie0}). For scenarios A and D
this is a much higher frequency than in scenarios B and C. Here
we clearly see that the transition energy $\omega$ is not the
most important parameter in the sense that it alone would set the
timescale for decoherence. Scenario D has a much shorter decoherence
time although it has almost the same $\omega$ as scenario A.
We already pointed out that an energy transfer (relaxation) from
system to environment or the other way around is not necessary
for decoherence \citep{Obada_2007}.

The timescale of decoherence is primarily given by the strength of
$\op H_{SE}$. This part of the Hamiltonian depends linearly on
the magnetic moments $g \mu$ of the environmental spins. In
Figure~\xref{fig:5} this parameter is chosen as $g \mu = 0.3359$ K/T, rising up to $g \mu
= 0.6717$ K/T in Fig.~\xref{fig:6} and to $g \mu = 1.3434$ K/T in
Fig.~\xref{fig:7}. We find that in all these cases scenario A
performs best regarding its decoherence time, but its advantage
becomes drastically smaller when the environment couples stronger to the system
(larger $g \mu$).

\clearpage

\begin{figure}[ht!]
\centering
\includegraphics*[clip,width=0.75\columnwidth]{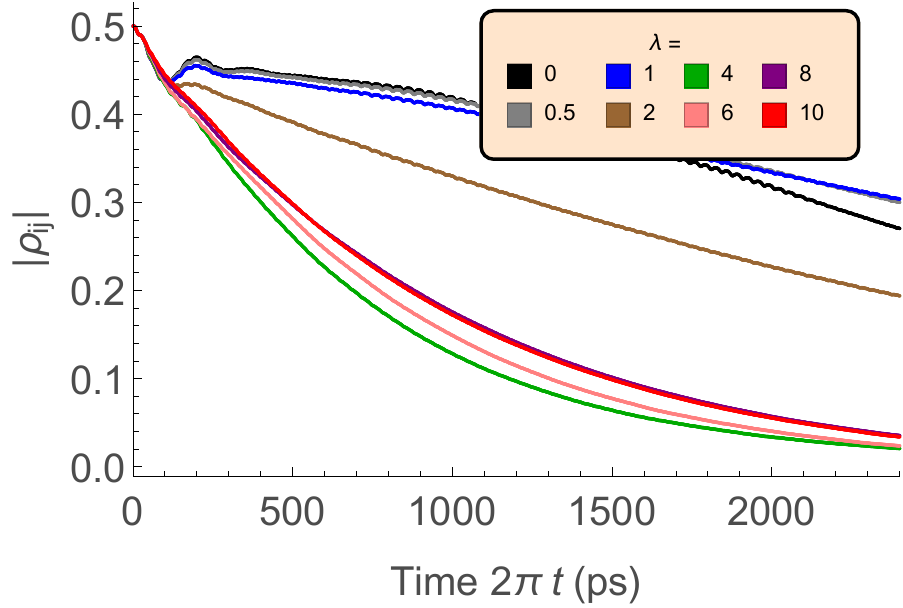}
\caption{\label{fig:11} Evolution of the absolute value of
  the relevant non-diagonal element $|\rho_{ij}|$ of the reduced
  density matrix \eqref{density} in initial scenario A for
  different scaling parameters $\lambda$ and fixed parameters  $g
  \mu = 0.3359$ K/T and $B = -1$ T. Increasing $\lambda$ leads
  to a transition from Gaussian to exponential decay law.}
\end{figure}

\begin{figure}[ht!]
\centering
\includegraphics*[clip,width=0.75\columnwidth]{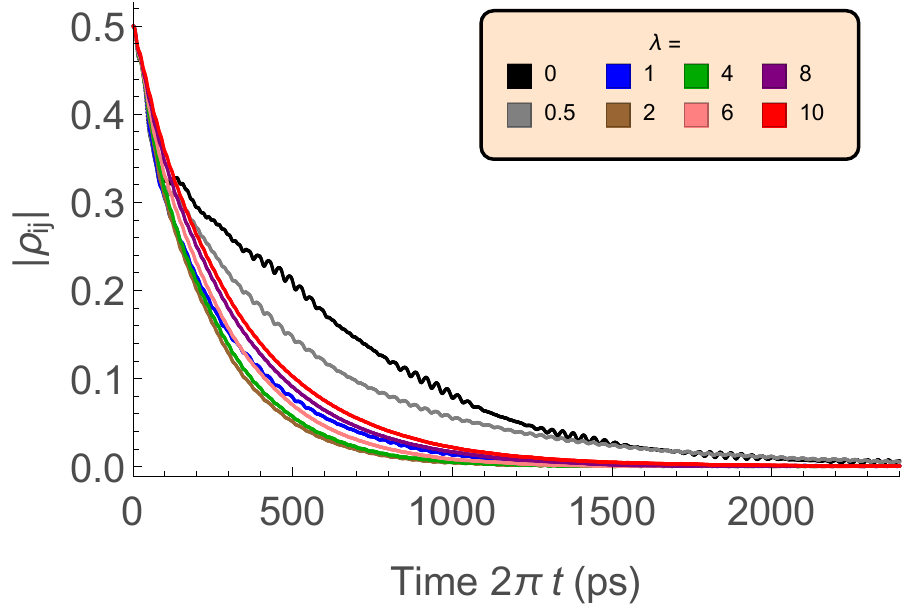}
\caption{\label{fig:12} Same as \figref{fig:11}, but with $g \mu = 0.6717$ K/T.}
\end{figure}

\begin{figure}[ht!]
\centering
\includegraphics*[clip,width=0.75\columnwidth]{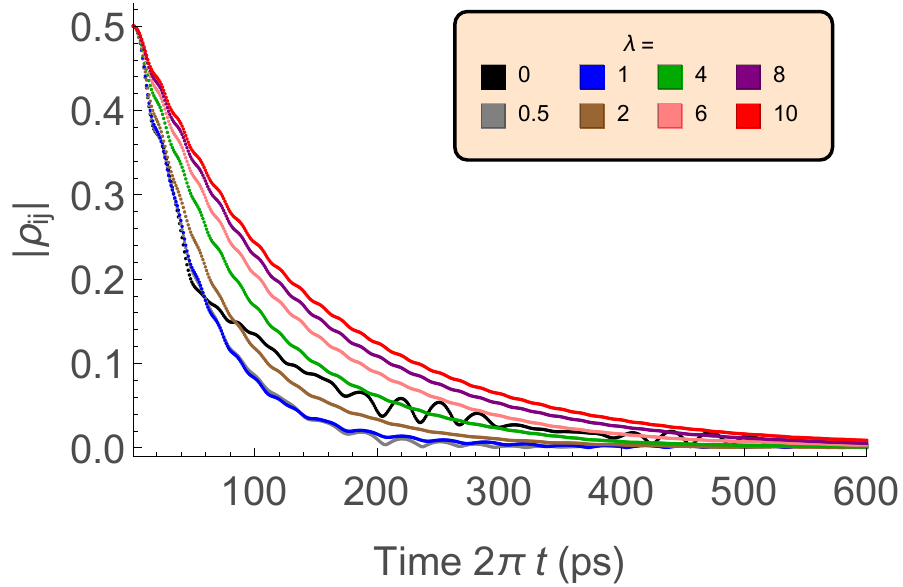}
\caption{\label{fig:13} Same as \figref{fig:11}, but with $g \mu
  = 1.3434$ K/T and a shorter period of time. In this regime a stronger interaction between
  environmental spins leads to a relative increase of decoherence
  time with growing $\lambda$.}
\end{figure}

Figures \xref{fig:11}, \xref{fig:12} and \xref{fig:13} show time
evolutions of initial scenario A for a lot of different values
of $\lambda$ and again different parameters $g \mu$. Here we see
that $\lambda$, which scales the strength of the
internal dynamic of the bath only, has a big impact on
the decoherence behavior of the system. In case of a weak coupling between
system and environment (\figref{fig:11}) a large value of
$\lambda$ changes the decoherence behavior from  approximately Gaussian to
exponential. Further increasing $\lambda$ leads to an oscillating decoherence time in
a certain range. 

For a strong coupling between system and environment
(\figref{fig:13}) the decay is always exponential even if
$\lambda = 0$, and in this regime the decoherence time can be
significantly improved by increasing $\lambda$.

Another effect we see in all three figures is that for a small
value of $\lambda \leq 1$ the decaying $|\rho_{ij}|$ has got a
superimposed oscillation. In the case of the Gaussian decay, this
oscillation is distinctive right at the beginning, while in the
exponential case it is visible at later times. 

\section{Summary}
\label{sec-5}

In our investigation we study the decoherence of a single 
Heisenberg-coupled spin dimer interacting with a spin bath.
We restrict ourselves to the effect of dipolar interactions on decoherence.
Other sources of decoherence such as phonons \citep{PrS:RPP00}
will be postponed to future investigations. We prepared the system
in different initial states (A, B, C, and D) and find that indeed
the clocklike superposition A has longer
decoherence times than other initial scenarios.
Our notion of decoherence time refers to a setting where a single
dimer subject to a spin bath is investigated. This differs from
usual investigations of decoherence, where $T_2^{*}$ times characterize
a dephasing ensemble.

The advantage of the clocklike scenario is impressive in a
regime of a small coupling between system and environment. The
difference between the scenarios is getting smaller with rising
strength of this coupling. In the case of a weak coupling between
system and environment the clock like superposition decays
Gaussian if the internal bath interactions are small enough,
otherwise the decay
becomes exponential and the decoherence time gets much worse.

In the case of a strong coupling between system and environment the
decay of superpositions is always
exponential, even if the environmental spins do not interact
with each other at all ($\lambda = 0$). One surprising result is that
in this regime the decoherence time can be improved
significantly by rising the internal interaction strength among
the bath spins. A possible explanation using Fermi's golden rule
might be that in this regime the smaller density of
bath states leads to slower decoherence as it does
equivalently for thermalization \cite{HKG:arXiv19,STS:PRB19}.
The effect of a small density of bath states on decoherence
needs to be further investigated, both theoretically
and experimentally.

A final remark concerns the special arrangement of the environmental
spins in our study. We studied several other arrangements, in particular
also one, where all environmental spins are situated in the lower hemispheres
around the dimer spins -- a situation that appears to be more adapted
to the experimental situation. However, our numerical experience yields that
the various geometries 
change the decoherence of the system primarily through a
different interaction strength between environmental spins (for
a fixed distance between dimer and bath), which is covered in
our model by choosing different factors $\lambda$.

\section*{Acknowledgment}

This work was supported by the Deutsche Forschungsgemeinschaft DFG
355031190 (FOR~2692); 397300368 (SCHN~615/25-1)). 
P.V. thanks Lennart Dabelow for fruitful discussions and Kristel
Michielsen, Hans De Raedt, as well as Fengping Jin for helpful
advise concerning advanced numerics.

%


\end{document}